\documentclass[12pt]{article}
\usepackage{amsmath}
\usepackage{times}
\usepackage{amssymb}
\usepackage{graphicx}
\usepackage{color}
\usepackage{multirow}
\usepackage[authoryear]{natbib}
\usepackage{rotating}
\usepackage{bbm}
\usepackage{latexsym}
\usepackage{booktabs}
\usepackage{textgreek}
\usepackage{tikz}
\usetikzlibrary{arrows.meta, positioning}

\usepackage{svg}
\usepackage{xcolor}
\makeatletter
\renewcommand{\fnum@figure}{\textbf{Figure~\thefigure}}
\makeatother

\newcommand{\captionfonts}{\normalsize}

\makeatletter  
\long\def\@makecaption#1#2{%
  \vskip\abovecaptionskip
  \sbox\@tempboxa{{\captionfonts #1: #2}}%
  \ifdim \wd\@tempboxa >\hsize
    {\captionfonts #1: #2\par}
  \else
    \hbox to\hsize{\hfil\box\@tempboxa\hfil}%
  \fi
  \vskip\belowcaptionskip}
\makeatother   

\begin{document}

{\noindent \bf \LARGE Inhibitory normalization of error signals improves learning in neural circuits}

\ \\
{ \large Roy Henha Eyono$^{ 1, 2, *}$, Daniel Levenstein$^{3}$, Arna Ghosh$^{4}$, Jonathan Cornford$^{5}$, Blake Richards $^{1,2,4,6,7,8}$} \\
{$^{1}$Mila-Quebec AI Institute}, 
{$^{2}$School of Computer Science, McGill University},
{$^{3}$Yale University},
{$^{4}$Google, Paradigms of Intelligence Team},
{$^{5}$Leeds University},
{$^{6}$Department of Neurology \& Neurosurgery, McGill University},
{$^{7}$Montreal Neurological Institute, McGill University},
{$^{8}$CIFAR Learning in Machines \& Brains Program}\\
\\
{$^{*}$ \text{Correspondence: \text{roy.eyono@mila.quebec}}}\\
%

{\noindent \bf Keywords:} Layer Normalization, Inhibition, Credit Assignment, Neural Networks

\thispagestyle{empty}
\markboth{}{Inhibitory Normalization of Error Signals}
%

\begin{center} {\bf Abstract} \end{center}

\noindent Normalization is a critical operation in neural circuits. In the brain, there is evidence that normalization is  implemented via inhibitory interneurons and allows neural populations to adjust to changes in the distribution of their inputs. In artificial neural networks (ANNs), normalization is used to improve learning in tasks that involve complex input distributions. However, it is unclear whether inhibition-mediated normalization in biological neural circuits also improves learning. Here, we explore this possibility using ANNs with separate excitatory and inhibitory populations trained on an image recognition task with variable luminosity. We find that inhibition-mediated normalization does not improve learning if normalization is applied only during inference. However, when this normalization is extended to include back-propagated errors, performance improves significantly. These results suggest that if inhibition-mediated normalization improves learning in the brain, it additionally requires the normalization of learning signals.

\section{Introduction}


Inhibitory plasticity has traditionally been studied as a means of maintaining excitation--inhibition (E--I) balance in neural circuits \citep{van1998chaotic, vogels2011inhibitory, deneve2016efficient}. In this work, we examine a complementary function of inhibitory circuits: their role in normalization, in which inhibitory neurons scale the activity of excitatory neurons relative to nearby neurons \citep{carandini2012normalization}.

There are several examples in which inhibitory interneurons facilitate normalization in the brain. For instance, \citet{atallah2012parvalbumin} showed that manipulating parvalbumin-positive interneurons in mouse V1 produces largely divisive and partially additive changes in pyramidal cell responses, suggesting that this class of interneurons can implement gain control consistent with normalization. Similarly, \citet{carandini2012normalization} described a feedforward normalization circuit in the fly antennal lobe, in which presynaptic local interneurons divisively scaled odor inputs.


Similar to the brain, normalization plays an important role in artificial neural networks (ANNs) \citep{wu2018group}. Layer normalization, which normalizes across units within the same layer \citep{ba2016layer}, has become a key component of transformer-based architectures and recurrent neural network models \citep{vaswani2017attention, ba2016layer}, and it leads to significant improvements in learning, especially for sequence modeling tasks \citep{xiong2020layer}.


While inhibitory circuits are critical for learning \citep{richards2010gabaergic}, it is unclear whether their importance for learning may relate to their role in normalization. This raises a question: Could normalization mediated by inhibitory circuits improve learning in the same way that layer normalization does in ANNs?


To understand the potential role of inhibitory normalization in learning, we trained ANNs with separate excitatory and inhibitory populations (EI-networks) on a visual classification task. First, using hard-coded layer normalization, we found that adding layer normalization to the EI-network significantly improved model training in the face of luminance changes.

We then asked whether layer normalization mediated by inhibitory circuits could provide a similar boost in performance. To answer this we trained the inhibitory cells in the network to perform layer normalization (I-normalization). We observed that, while I-normalization could successfully center and scale neural activity, it did not produce the same benefits for learning as the hard-coded layer normalization operation. Closer examination of the layer normalization operation revealed that its primary contribution to learning lay not in stabilizing activations, but in shaping gradients during backpropagation \citep{xu2019understanding}. 

This insight led us to implement an additional lateral inhibitory mechanism to normalize the back-propagated error signals in EI-networks. With this form of I-normalization, the EI-network was able to recapitulate the performance benefits of hard-coded layer normalization.


Altogether, our results support the idea that inhibition-mediated normalization could be one of the reasons that inhibition is important for learning in real neural networks. But, our results also suggest that normalizing activity alone would not be sufficient. To obtain the benefits for learning, I-normalization would need to not only normalize the neural activity, but also any signals used for learning. This has interesting implications for inhibitory circuits in the brain that target the apical dendrites of pyramidal neurons, where learning signals may be received.

\section{Results}

\subsection{Layer-normalization improves perceptual invariance}

\begin{figure}[!t]
    \makebox[\linewidth]{%
    \includegraphics[width=1.0\textwidth]{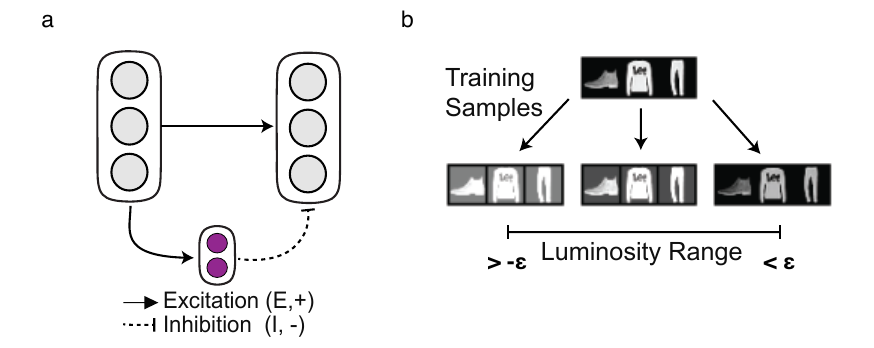}
  }
  \vspace{-20pt}
    \caption{
        \textbf{Schematic of the Excitatory-Inhibitory (EI) network with Layer Normalization and the perceptual invariance task.}
        \textbf{a:} Feedforward EI network architecture. Gray units represents the \textit{Excitatory (E) population}, and purple unit represents the \textit{Inhibitory (I) population}. Outgoing synaptic weights share the same sign ($\text{E}, +$ or $\text{I}, -$).
        \textbf{b:}  To test perceptual invariance, a shift is applied to each individual image in the FashionMNIST dataset during both training and testing. For every image, a unique constant $\Delta$ is sampled within the threshold $|\Delta| < \epsilon$. The figure displays three example augmentations to illustrate how the shift varies across the allowable range. \newline
    } 
    \label{fig:fig0}
\end{figure}

To study how I-normalization could impact learning, we used an ANN that enforces Dale’s principle by constraining each neuron to be either purely excitatory or purely inhibitory, so that all outgoing synaptic weights from the same neurons share the same sign (EI-networks; Fig. \ref{fig:fig0}a). Following prior work \citep{cornford2021learning}, such networks have been shown to achieve performance comparable to standard ANNs when trained with gradient descent. In these networks, the activity of excitatory units at layer $\ell$ is governed by the interaction between a direct excitatory drive and an indirect, feed-forward inhibitory pathway. The activity in the network is calculated as follows:

\begin{align*}
    \mathbf{h}_{\ell}^I &= \mathbf{W}_{\ell}^{IE} \mathbf{h}_{\ell-1}^E, \\
    \mathbf{z}_{\ell}   &= \mathbf{W}_{\ell}^{EE} \mathbf{h}_{\ell-1}^E - \mathbf{W}_{\ell}^{EI} \mathbf{h}_{\ell}^I, \\
    \mathbf{h}_{\ell}^E &= \phi(\mathbf{z}_{\ell} + \mathbf{b}_{\ell}),
\end{align*}

\noindent where $\mathbf{h}_{\ell}^E$ and $\mathbf{h}_{\ell}^I$ represent the activity vectors of the \textit{excitatory} and \textit{inhibitory} populations at layer $\ell$, respectively. The weight matrices are defined as follows: $\mathbf{W}_{\ell}^{EE}$ is the direct excitatory-to-excitatory connection, $\mathbf{W}_{\ell}^{IE}$ projects activity from the previous excitatory layer to the current inhibitory population (feedforward inhibition), and $\mathbf{W}_{\ell}^{EI}$ represents the inhibitory weights that subtractively modulate the excitatory drive. The term $\mathbf{b}_{\ell}$ denotes the learnable bias vector, and $\phi(x) = \text{ReLU}(x)$ is the non-linear activation function applied element-wise.

We trained the EI-network on a modified Fashion MNIST categorization task with shifts in luminosity. We did this because we reasoned that normalization would be especially important for handling input distribution variability. Specifically, for each image, we created a series of augmentations of luminance by translating the pixel values by a constant, $\Delta$, sampled within the threshold range $\Delta \sim \text{Uniform}(-\epsilon,+\epsilon)$. This was done for every sampled image during both training and test time (see Methods). The variable $\epsilon$ then served as a hyperparameter to adjust the magnitude of the range of luminance shifts in the data distribution (Fig. \ref{fig:fig0}b). Succeeding in this task requires perceptual invariance to changes in luminosity, a capability that animals routinely exhibit and which is critical for navigating a dynamic environment.

\newpage

\begin{figure}[!t]
\centering
    \makebox[\linewidth]{%
    \includegraphics[width=1.0\textwidth]{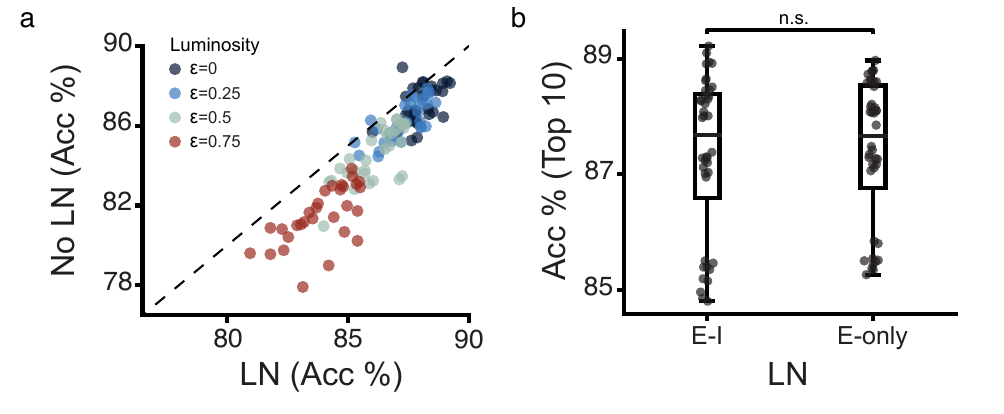}
  }
  \vspace{-20pt}
    \caption{
        \textbf{Layer normalization (LN) improves perceptual invariance in Excitatory-Inhibitory (EI) networks.}
        \textbf{a:} Test accuracy (Acc \%) comparison of EI networks \textit{with} LN (x-axis) to those \textit{without} LN (y-axis). Data points represent performance across 30 hyperparameter combinations (layer widths and E/I learning rates) and four luminosity ranges ($\epsilon = 0, 0.25, 0.5, 0.75$). Points below the dashed diagonal line indicate cases where networks \textit{with} LN performed better.
        \textbf{b:} Top-10 test accuracy comparison of an EI network with LN against an Excitatory-only (E-only) network, also with LN. The box plots summarize the distribution across the same 30 hyperparameter combinations reported in panel $\mathbf{a}$. \newline 
    }
    \label{fig:fig1}
\end{figure}

We first examined the impact of hard-coded layer normalization in these EI-networks (Fig. \ref{fig:fig0}a). Specifically, we applied a centering and scaling operation to the excitatory pre-activations in each layer in order to bring the pre-activation mean to zero and variance, one:

\begin{align*}
    h_{\ell}^E     &= \phi(\hat{z}_{\ell}), \\
    \hat{z}_{\ell} &= \frac{z_{\ell}-\mu_{\ell}}{\sqrt{\sigma_{\ell}^2 + c}}, \\
    \mu_{\ell}     &= \frac{1}{n_{\ell}}\sum_{i=1}^{n_{\ell}}{z_{\ell}^i}, \\
    \sigma_{\ell}  &= \frac{1}{n_{\ell}}\sum_{i=1}^{n_{\ell}}{(z_{\ell}^i - \mu_{\ell})^2},
\end{align*}

\noindent where $n_{\ell}$ is the number of excitatory neurons in layer $\ell$, $c = 1 \times 10^{-5}$ is a small constant to prevent division by zero, and the bias term has been omitted for simplicity. We found that this layer normalization operation improved training, with the improvement being more pronounced for larger ranges of luminance shifts (Fig. \ref{fig:fig1}a).

To assess the empirical contribution of the inhibitory units, we selectively ablated them before training. In this condition, only the excitatory weights were trained, with layer normalization alone preventing activity blow-up. Despite having fewer parameters than their EI-network counterparts, the E-networks with layer normalization showed no statistically significant difference in performance to the EI-network with layer normalization (Fig. \ref{fig:fig1}b), suggesting that, for this task, training the inhibitory units on the task loss (cross-entropy) provides little to no benefit when hard-coded layer normalization is present. In other words, inhibition does not meaningfully contribute to task performance under these conditions, suggesting that the inhibitory units’ capacity could instead be directed toward implementing layer normalization.

\subsection{Learned inhibition normalizes excitatory activity}

\begin{figure}[t!]
    \makebox[\linewidth]{%
    \centering
    \includegraphics[width=1.0\textwidth]{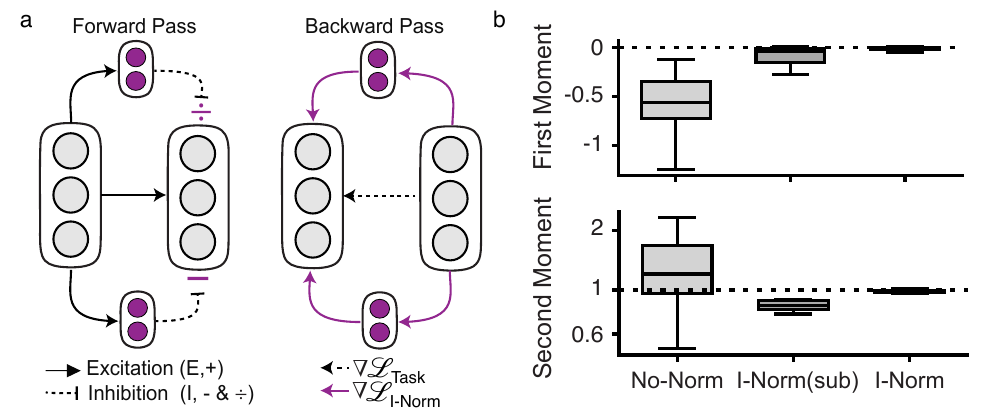}
  }
  \vspace{-20pt}
    \caption{
        \textbf{Inhibitory populations learn to implement layer normalization of excitatory activity.}
        \textbf{a:} Schematic showing how the inhibitory circuit (purple) is trained locally via the $\mathcal{L}_{\text{I-Norm}}$ loss (purple lines) to normalize excitatory activity. Excitatory to excitatory weights are updated only by the task loss $\mathcal{L}_{\text{Task}}$ (dotted left arrow). Forward Pass: Inhibition performs subtractive ($-$) and divisive ($\div$) modulation. Backward Pass: Inhibitory gradients enforce layer-normalized excitatory statistics.
        \textbf{b:} Box-and-whisker plots of the first and second moments of excitatory activations. Each plot compares three conditions: No-Norm, Subtractive-only I-Norm (sub), and I-Norm (as depicted in a). Each box plot summarizes model results aggregated across the sampled range of $\epsilon$ luminosity augmentations.\newline
    }

    \label{fig:fig2}
\end{figure}


We next asked whether hard-coded layer normalization could be removed entirely and replaced with layer normalization implemented by inhibitory neurons. To implement layer normalization with inhibitory neurons, we used two distinct inhibitory populations: one providing subtractive inhibition and the other providing divisive inhibition. This design accounts for the functional diversity of inhibitory subtypes, which can implement divisive or subtractive operations or both depending on the context \citep{wilson2012division, pouille2013contribution, el2014response}. We will refer to these networks as I-normalization (or I-Norm) networks. Their activity was calculated as follows:

\begin{align}
\label{eq:inorm}
    \nonumber h_{\ell}^E &= \phi(z_{\ell}) \\
    z_{\ell}   &= \frac{W_{\ell}^{EE} h_{\ell-1}^E - W_{\ell-1}^{EI} h_{\ell}^{I}}{\sqrt{U_{\ell}^{EI} h_{\ell}^D}},
\end{align}

where $h_{\ell}^D = U_{\ell}^{IE} h_{\ell-1}^E$ represents the divisive inhibition population, $h_{\ell}^I = W_{\ell}^{IE} h_{\ell-1}^E$ represents the subtractive inhibition population, and $U_{\ell}^X, W_{\ell}^X$ represents the synaptic weights associated with each respective inhibitory population.

We then introduced a new normalization loss ($\mathcal{L}_{\text{I-Norm}}$), which was applied only to the inhibitory pathway weights ($W^{EI}$, $W^{IE}$, $U^{EI}$, $U^{IE}$), while the excitatory weights ($W^{EE}$) were trained solely with the cross-entropy loss from the Fashion MNIST categorization task ($\mathcal{L}_{\text{Task}}$, see Methods). The normalization loss was calculated as:

\begin{equation*}
    \mathcal{L}_{I-Norm} = \left ( \frac{1}{n} \sum_{i=1}^n h_i^E\right ) ^2 + \left ( \frac{1}{n} \sum_{i=1}^n {(h_i^E)}^2 -1 \right )^2.
\end{equation*}

This loss was designed to optimize the first and second moments of the excitatory unit activity ($h^E$). Figure \ref{fig:fig2}a provides a schematic of this setup, showing the distinct inhibitory pathways and how the loss gradients propagate through them.

Analysis of excitatory unit activity revealed that the inhibitory circuits effectively learned to normalize excitatory activity (Fig. \ref{fig:fig2}b). In contrast, networks without layer normalization exhibited first and second moments that were off the target ($\mu=0, \sigma^2=1$). Figure \ref{fig:fig2}b similarly shows that a purely subtractive inhibitory circuit does not match the layer normalization target statistics as effectively as the segregated inhibitory approach depicted in Figure \ref{fig:fig2}a, despite being trained with the same I-norm loss objective.

These results demonstrate that inhibitory circuits can effectively recapitulate the impact of hard-coded layer normalization on neural activity. Importantly, unlike hard-coded layer normalization, which requires a population-level computation to aggregate statistics, the learned I-Normalization depends only on the feedforward activity of the preceding layer ($h_{\ell-1}^E$, see Eq. \ref{eq:inorm}). This makes normalization inherently ``predictive": the inhibitory population must anticipate excitatory activity, rather than simply enforcing a hard-coded normalization operation.

\subsection{Normalizing error signals recapitulates layer normalization function}

\begin{figure}[!t]
    \makebox[\linewidth]{%
    \includegraphics[width=1.0\textwidth]{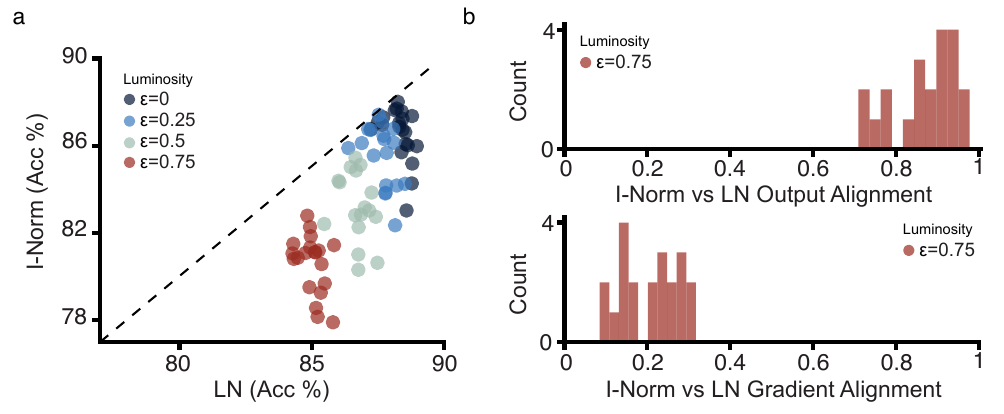}
  }
  \vspace{-20pt}
    \caption{
        \textbf{I-Norm networks struggle to recapitulate the performance of LN in EI networks.}
        \newline
        \textbf{a:} Test accuracy comparison between LN (x-axis) and I-Norm (y-axis) across hyperparameter and luminosity ranges ($\epsilon = 0, 0.25, 0.5, 0.75$). Each point represents a single network training run. Points falling below the dashed diagonal line indicate cases where LN achieved higher test accuracy than I-Norm.
        \textbf{b:} Layer-wise alignment between I-Norm and LN. We quantify alignment as the cosine similarity between I-Norm and LayerNorm (LN) across all network layers for outputs (top) and gradients (bottom). All results correspond to the highest luminosity range ($\epsilon = 0.75$).
    }
    \label{fig:fig3}
\end{figure}

We next examined the performance of the trained I-Norm networks on the perceptual invariance task. Although the I-norm network normalized excitatory activity, it still underperformed compared to EI-networks with hard-coded layer normalization (Fig. \ref{fig:fig3}a). These observations point to a key mechanism found in layer normalization being absent from I-Norm networks.


One possibility, motivated by observations in the machine learning literature \citep{xu2019understanding}, is that the effectiveness of hard-coded layer normalization depends on how it transforms error signals during error propagation, rather than how it shapes activity during forward processing. To test this idea, we compared the activity vectors and weight updates during training in I-Norm networks with the a version of the network with inhibition removed and hard-coded layer normalization. Despite the strong cosine similarity in the activity statistics (Fig. \ref{fig:fig3}b, \textit{top}), the weight updates between the two networks were poorly aligned (Fig. \ref{fig:fig3}b, \textit{bottom}). This suggests that the impact of layer normalization is related to its impact on the weight updates, rather than its impact on forward-pass activity, per se.

\begin{figure}[!t]
    \makebox[\linewidth]{%
    \centering
    \includegraphics[width=1.0\textwidth]{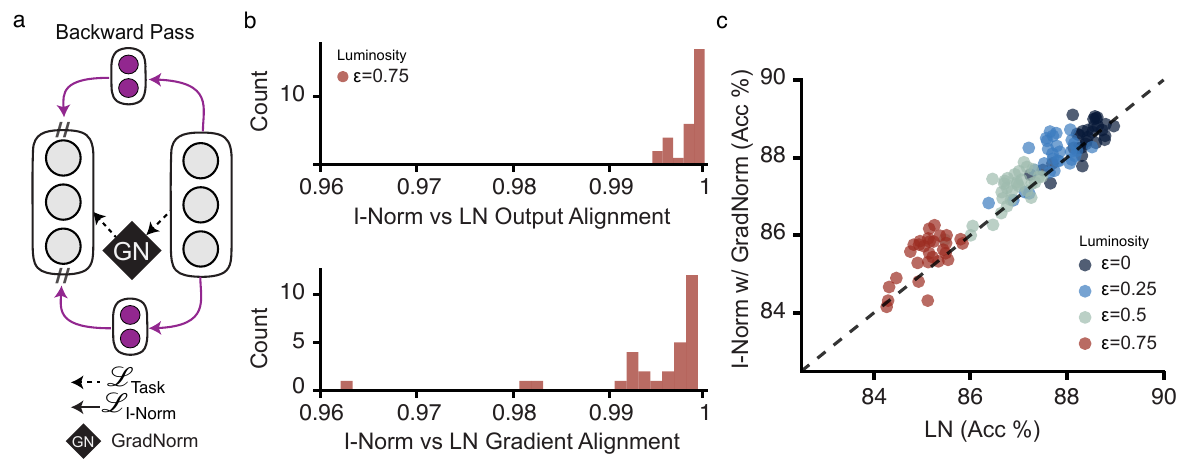}
  }
  \vspace{-20pt}
    \caption{
        \textbf{Hard-coded LN gradients in I-Norm networks restore LN performance in EI networks.}
        \textbf{a:} Schematic illustrating the Backward Pass of the I-Norm network incorporating, \textit{GradNorm}, from equation \ref{eqn: ln_der}. The $\text{GradNorm}$ operation is applied to the backward signal ($\delta$) to enforce the LN gradient.
        \textbf{b:} Average cosine similarity across all network layers. The top and bottom panels show the similarity between  LN and $\text{I-Norm (\textit{with GradNorm})}$ for outputs and gradients, respectively.
        \textbf{c:} Test accuracy (Acc \%) comparison. LN network performance (x-axis) versus I-Norm network with $\text{GradNorm}$ (y-axis). Data is shown across 30 hyperparameter initializations and four luminosity ranges ($\epsilon = 0, 0.25, 0.5, 0.75$). Points clustered along the dashed diagonal line indicate a strong match in performance between the two models.
    }
    \label{fig:fig4}
\end{figure}

Formal analysis of the weight updates confirmed that the effect of hard-coded layer normalization on learning arises from its influence on the weight updates themselves (see Appendix, 5.1). Specifically, for a network with layer normalization, the partial derivative used to update weights, $\frac{\partial \mathcal{L}_{Task}}{\partial z_i}$, corresponds to a normalized version of the propagated error signal, which we denote as $\hat{\delta}_i$ for neuron $i$. This relationship can be expressed as follows (see Appendix for a full derivation):

\begin{align}
\nonumber \Delta W^{EE} &\propto -\frac{\partial \mathcal{L}_{Task}}{\partial z} = -\hat{\delta} \\
\hat{\delta}_i &= 
\underbrace{\textcolor{purple}{\frac{1}{\sqrt{\sigma^2 + c}}}}_{\makebox[0pt][c]{\textcolor{purple}{\normalsize scale}}}
\Bigg(
\delta_i 
- \underbrace{\textcolor{orange!70!black}{\frac{1}{n} \sum_{j=1}^{n} \delta_j}}_{\makebox[0pt][c]{\textcolor{orange!70!black}{\normalsize center}}}
- \underbrace{\textcolor{teal!60!black}{\frac{\hat{z}_i}{n} \sum_{j=1}^{n} \delta_j \cdot \hat{z}_j}}_{\makebox[0pt][c]{\textcolor{teal!60!black}{\normalsize decorrelate}}}
\Bigg),
\label{eqn: ln_der}
\end{align}

where $z_i$ is the activation of unit $i$, $\hat{z}_i$ is its normalized activation, $\delta_i$ is the backpropagated error before normalization, $\hat{\delta}_i$ is the error after normalization, $\mu$ and $\sigma^2$ are the mean and variance across the $n$ units, and $c$ is a small constant for numerical stability. For clarity, we have omitted the layer index $\ell$ and color-coded the terms to highlight their different functional roles. As shown in the equation, layer normalization transforms the propagated error signals in three distinct ways:
(1) it rescales the errors according to the variance of the excitatory activity;
(2) it centers the errors by removing their mean;
and
(3) it orthogonalizes the error signal from the activations.


We next evaluated I-Norm networks using the gradient modification in Equation \ref{eqn: ln_der}, which we call GradNorm. In this setting, inhibition handles activity normalization, while the error calculations are hard-coded (Eqn. \ref{eqn: ln_der}; Fig. \ref{fig:fig4}a). GradNorm significantly improved the alignment of I-Norm weight updates with those of standard layer normalization (Fig. \ref{fig:fig4}b), successfully recapitulating the performance gains associated with hard-coded layer normalization (Fig. \ref{fig:fig4}c). This performance held across all ranges of luminosity, despite the lack of perfect output alignment(Fig. \ref{fig:fig4}b).


Together, these analyses and experiments indicate that the performance improvements observed in EI-networks with layer normalization derive from its effect on error signal propagation. This suggests that achieving the benefits of layer normalization in I-Norm networks would potentially require an additional inhibitory population capable of normalizing error signals.


\subsection{Mean centering of error signals is the most salient component of credit assignment normalization}

\begin{figure}[!t]
\makebox[\linewidth]{%
    \includegraphics[width=1.0\textwidth]{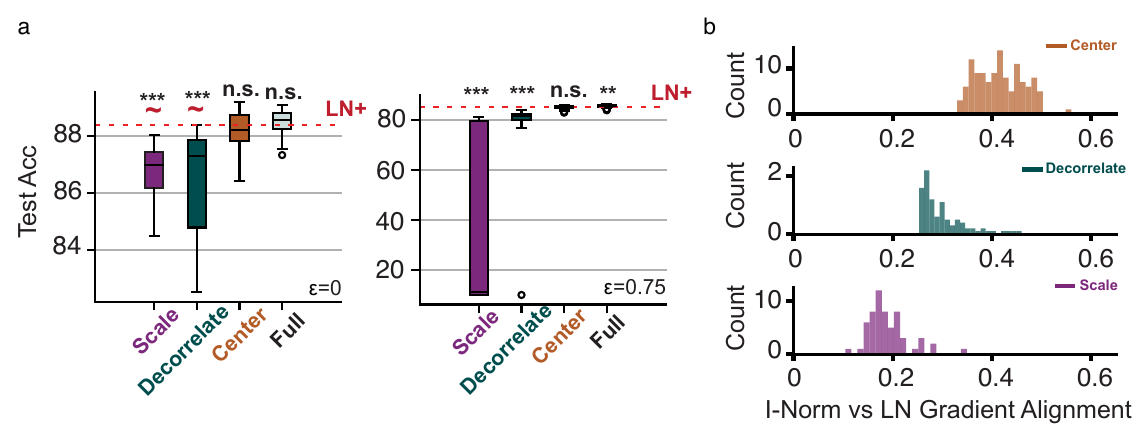}
  }
  \vspace{-20pt}
    \caption{
        \textbf{Gradient centering, not scaling or decorrelating, drives LN performance recovery in I-Norm networks.}
\textbf{a:} Box plots of test accuracy comparing \textit{Scale}, \textit{Decorrelate}, \textit{Center}, and \textit{Full} LN gradient components applied to the I-Norm network. The red dashed line ($\mathbf{LN^+}$) shows the baseline performance of the hard-coded LN network. Results are separated by luminosity range ($\epsilon = 0, 0.75$). Significance is indicated comparing components to the $\text{LN}^+$ baseline. Red tildes ($\sim$) denote outliers performing near random chance, omitted to preserve plot scaling.
\textbf{b:} The panels contrast the LN gradient with specific gradient components trained on an I-Norm network: \textit{Center} (top, orange), \textit{Decorrelate} (middle, teal), and \textit{Scale} (bottom, purple).\newline 
    }
    \label{fig:fig5}
\end{figure}


To better understand which aspects of gradient normalization drive performance, we first examined the contribution of each term in the layer-normalization gradient equation. In particular, we asked whether the performance of I-Norm networks with hard-coded GradNorm hinges more on the \textbf{\textcolor{purple}{scaling}}, \textbf{\textcolor{orange!70!black}{centering}}, or \textbf{\textcolor{teal!60!black}{decorrelation}} of the error signals (Eqn. \ref{eqn: ln_der}).


To investigate this, we trained I-Norm networks with hard-coded modifications to the gradient calculations (Eqn. \ref{eqn: ln_der}), applying each of the three operations in isolation. Across luminance levels, we found that centering was critical for performance, while scaling and decorrelation had little impact. Networks trained with centering alone achieved test accuracies comparable to those of hard-coded layer normalization at luminance ranges of 0 and 0.75 (Mann-Whitney $U$ test, $p = 0.4641$ and $p = 0.6099$, respectively), whereas networks trained with only scaling or decorrelation performed significantly worse across luminance levels (Fig. \ref{fig:fig5}a).


We further confirmed these findings by analyzing the alignment of the gradients in I-Norm networks with those trained using standard layer normalization. Networks trained with the centering operation showed higher alignment with the layer normalization networks compared to networks trained with either scaling or decorrelation alone (Fig. \ref{fig:fig5}b).


These results indicate that the centering of gradient calculations induced by layer normalization is the key factor driving its impact on training for the task. Motivated by this, we next aimed to implement the centering operation using an inhibitory circuit.

\subsection{Centering of credit assignment signals via lateral inhibition}


An inhibitory implementation of the mean-centering operation must (i) pool information across the $\delta_i$'s within the layer, and
(ii) broadcast back a common signal that each excitatory neuron can subtract
from its update. But, inhibitory units in real neural circuits cannot all possess the same exact synaptic weights. This raises a
fundamental question: \emph{can random inhibitory synapses implement
gradient centering?}

Several theories argue for gradient approximation in the brain, each sharing a common requirement: error signals (i.e., $\delta_i$) are propagated between regions or layers \citep{richards2019deep, guerguiev2017towards, lillicrap2016random, whittington2017approximation, lillicrap2020backpropagation}. Motivated by this, we assume here for sake of theorizing that such across-layer error signals are available, and focus on a more specific question: given access to the $\delta_i$ within a layer, how could an inhibitory circuit implement the mean-centering operation?

\begin{figure}[!t]
\makebox[\linewidth]{%
    \includegraphics[width=1.0\textwidth]{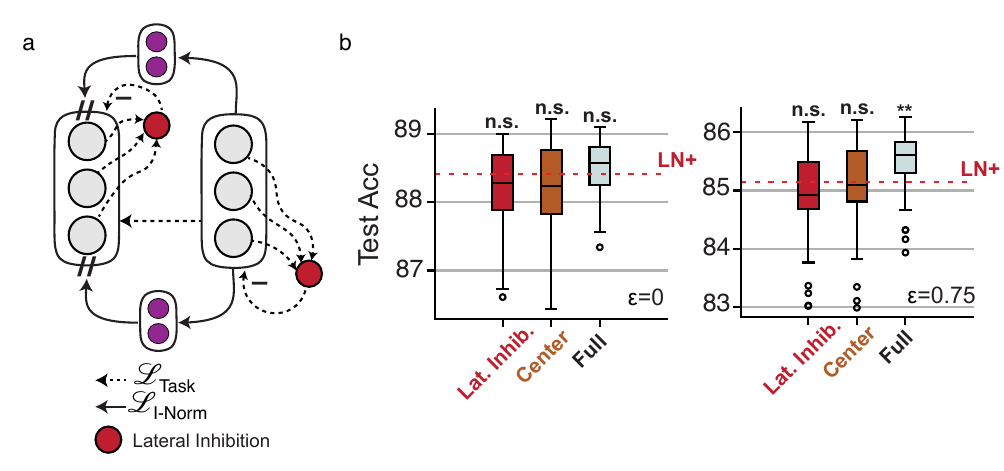}
  }
  \vspace{-20pt}
    \caption{
        \textbf{Lateral inhibition with fixed synaptic weights implements gradient centering.}
\textbf{a:} Schematic of the I-Norm network with a lateral inhibition pathway. The inhibitory unit (red) pools and transforms excitatory activity (gray) using fixed, random connections to compute the centering term (mean).
\textbf{b:} Test accuracies (Acc \%) of the lateral inhibition solution, explicit gradient centering, and the full layer normalization gradient applied to the I-Norm network. The red dashed line ($\mathbf{LN^+}$) shows the baseline performance of the explicit layer normalization network. Results are averaged across hyperparameter and luminosity ranges ($\epsilon = 0,  0.75$). Significance is shown relative to the LN+ baseline. \newline
    }
    \label{fig:fig6}
\end{figure}

Below, we establish a theoretical guarantee showing that a single inhibitory
unit with fixed, random synaptic weights can indeed approximate the population
mean of error signals.

\noindent \textbf{Theorem 1: Mean estimation via fixed random inhibition}

Let $\{\omega_i\}_{i=1}^{n}$ be i.i.d. positive random variables with $\mathbb{E}[\omega_i] = \mu > 0$ and $\operatorname{Var}(\omega_i) < \infty$ that parameterize a set of $n$ synaptic weights, $\{\nu_{i,n}\}_{i=1}^{n}$, whose values sum to $1$.Let $\{\delta_i\}_{i=1}^{n}$ be a bounded sequence of error signals with an empirical mean denoted by $\bar{\delta}$, i.e.,$$\bar{\delta} = \frac{1}{n}\sum_{i=1}^n \delta_i.$$

Define the normalized synaptic weight onto inhibitory neuron, $i$ as $$\nu_{i,n} = \frac{\omega_i}{\sum_{j=1}^n \omega_j}.$$
If we consider the inhibitory error pooling operation, which can be expressed as the dot product between the $n$-dimensional weight vector $\boldsymbol{\nu}_n = [\nu_{1,n}, \dots, \nu_{n,n}]^\top$ and the error vector $\boldsymbol{\delta}_n = [\delta_1, \dots, \delta_n]^\top$:$$s_n = \boldsymbol{\nu}_n \cdot \boldsymbol{\delta}_n = \sum_{i=1}^n \nu_{i,n}\,\delta_i.$$
Then, in the limit as $n \to \infty$, the pooled signal converges to the empirical mean:$$\lim_{n \to \infty} s_n = \bar{\delta} \quad\text{almost surely}.$$

\hfill\textit{(End of Theorem 1)}

The proof of this theorem is provided in the Appendix. Theorem 1 shows that pooling errors via normalized random synapses becomes equivalent to uniform averaging in the large-$n$ limit. Subtracting this inhibitory signal, therefore, could approximate gradient centering. Motivated by this theoretical guarantee, we construct a simple lateral inhibitory circuit (Fig.~\ref{fig:fig6}a), inspired by ``blanket'' inhibition in cortex, in which a single inhibitory pool targets large populations of excitatory cells \citep{karnani2014blanket}.


We tested this lateral inhibition mechanism for error normalization and compared its performance to networks using explicit centering or full gradient normalization. Across all luminance ranges, test accuracy with lateral inhibition was comparable to the explicit centering and full gradient-normalization baselines, with no significant differences found at luminance range = 0 (Mann-Whitney $U$ test, $p = 0.4641$) or luminance range = 0.75 (Mann-Whitney $U$ test, $p = 0.6099$) (Fig. \ref{fig:fig6}b).


Together, these results show that the learning benefits of layer normalization can be reproduced using three distinct inhibitory populations: two inhibitory populations that estimate the mean and variance to normalize excitatory activity in the forward pass and another lateral inhibitory population that centers error signals to normalize gradient updates.


\section{Discussion}

Layer normalization enhances learning in ANNs by stabilizing activations in the forward pass and regularizing error signals in the backward pass. In our work, we leveraged a neural network with distinct excitatory and inhibitory circuits to understand the potential role of inhibitory normalization in learning in neural circuits. Using local objectives, the inhibitory circuits learned to stabilize excitatory activations, but lacked the additional learning benefits provided by standard layer normalization. Through ablation studies, we showed that centering the gradient alone is sufficient to recover these benefits, even without explicitly regularizing error signals. Moreover, we demonstrated that such centering can be achieved through lateral inhibitory circuits with fixed random weights, leading to performance improvements approaching that of standard layer normalization. Altogether, our results suggest that if inhibitory normalization in the brain helps learning it may be due to there being multiple inhibitory populations with distinct roles, some related to normalizing activity, and others related to normalizing error signals used for learning.

Previous studies have emphasized how inhibition can center (subtractive) or scale (divisive) neuronal activity: Somatic-targeting PV interneurons typically scale down (divide) responses, whereas dendrite-targeting SST interneurons subtract from them \citep{wilson2012division, pouille2013contribution} . Consistent with this, \cite{atallah2012parvalbumin} showed that increasing PV-cell activity produces a linear combination of additive and multiplicative changes in pyramidal firing, effectively a form of gain control with bias while maintaining tuning specificity. These findings collectively support the idea that PV-cells could implement activity normalization. Our work builds on this foundation by training inhibitory cells to perform normalization on excitatory activity, demonstrating that such circuits can recapitulate both the centering and gain control functions of standard layer normalization.

However, previous models of inhibitory normalization have focused primarily on regulating neuronal activity. Learned divisive normalization frameworks \citep{burg2021learning, shen2021correspondence}, for example, describe how cortical circuits can perform normalization but do not consider how inhibitory mechanisms might also regulate error signals. Our findings address this gap by showing that forward-pass normalization by inhibition alone cannot reproduce the full learning benefits of layer normalization. We propose that inhibitory subcircuits, potentially SST-subtypes targeting apical dendrites where error-related signals arrive or neurogliaform cells \citep{overstreet2015neurogliaform}, can normalize credit assignment signals and thereby support efficient and stable learning. This aligns with recent theories of burst-dependent backpropagation \citep{payeur2021burst}, which suggest that inhibitory microcircuits are essential in shaping back-propagated error signals rather than merely transforming neural responses. In this view, inhibition not only stabilizes neuronal activity but also contributes to modulating credit assignment signals, revealing a hypothetical, previously unrecognized role for inhibition.

One of the more interesting insights from our work is the ability of inhibitory circuits to infer the mean and variance of downstream excitatory activity from an earlier layer using a simple regularization loss, without direct knowledge of the excitatory activations. In the context of ANNs, predicting these statistics eliminates the need for hard-coded layer normalization. Although this mechanism is not present in standard networks, it could help maintain stable representations under varying sensory conditions, such as large shifts in luminance for image categorization or other distributional changes, reducing the need for explicit gain, bias, or error correction in variance computation.

Reflecting on the role of gradient normalization, a particularly striking result of our study was that centering the gradients alone was sufficient in recapitulating layer normalization's performance on the task, indicating that the mean component of the gradient carries the majority of the functional impact. In practice, the mean computation occasionally reverses the signs of individual gradients, which in the context of gradient descent may produce substantially different learning dynamics. Why such sign changes can improve learning remains unclear and potentially represents an interesting direction for future investigation.

Here, we implemented the gradient-centering operation using a lateral inhibitory circuit with fixed random weights. This mechanism is conceptually related to feedback alignment \citep{lillicrap2016random}, in which fixed random feedback weights transmit useful gradient signals. In our case, the fixed random inhibitory weights serve to compute the mean of the gradient rather than the exact backpropagated values. This observation raises further questions for future research, including whether using random inhibitory weights to maintain homeostasis could represent a biologically plausible alternative to approximating precise gradient signals.

All things considered, this work has some important limitations that should be considered. First, our empirical findings rely primarily on a perceptual invariance task in which the dominant inductive bias is a global shift in pixel intensities. In this task, mean-centering would naturally be the most functionally relevant component of normalization. Consequently, subtracting the mean from the activations directly targets the structure of the perturbation. In more complex sensory domains, there is no guarantee that the mean will remain the most salient statistic to normalize. Natural images, for example, exhibit variance fluctuations, which might require inhibitory circuits to compensate beyond centering the gradient. Although centering the gradient works well for our perceptual invariance, we are yet to test whether it generalizes to other tasks.

Finally, a core conceptual limitation is that the random-weight inhibitory mechanism we employ to center error signals is not biologically grounded; for example, it is linear and and operates under the assumption that the random weights $\nu_i$ are unit vectors. Nonetheless, it provides a high-level illustration of how lateral inhibition with random synaptic connections could implement normalization of error signals. Thus, the biological claims of this work are best interpreted as a high-level algorithmic hypothesis, rather than as a proposal for a physiological mechanism.

Future work should evaluate inhibitory normalization in tasks where additional statistics beyond the mean (e.g., variance, covariance, or sparsity) are behaviorally relevant. Furthermore, additional studies could build biophysical models that examine the relationship between plasticity rules and normalization in more realistic inhibitory circuits. By expanding both the task domain and the realism of the circuit motifs studied, future work may uncover a more complete and unified account of how normalization driven by inhibition could impact learning in neural circuits.






\section{Methods}
\label{sec:methods}

\subsection{Dataset}

We constructed a modified version of the Fashion-MNIST classification task that incorporates shifts in image luminosity. All images were normalized to lie within the range [0, 1]. For each image, we generated luminance augmentations by adding a constant offset, $\Delta$, to all pixel values, where $\Delta \sim \text{Uniform}(-\epsilon, +\epsilon)$. Since the pixel values are bounded between 0 and 1, the augmented images were clamped to remain within this range.

The parameter $\epsilon$ served as a hyperparameter controlling the magnitude of the luminance shifts. We conducted experiments with $\epsilon \in {0, 0.25, 0.5, 0.75}$, where $\epsilon=0$ corresponds to the standard Fashion-MNIST dataset.

Each model was trained on 60,000 images and evaluated on a test set of 10,000 images, both of which included luminosity-shifted samples. Reported accuracies in the manuscript refer to model performance on the test dataset.

\subsection{Layer Normalization}

Layer Normalization (LN) \citep{ba2016layer} normalizes the pre-activations of a layer across hidden units for each input sample, rather than across the batch.

Given a vector of pre-activation inputs $\mathbf{z} = (z_1, \dots, z_N)$ to a layer of $N$ hidden units, LN computes the mean and variance over the hidden units for a single sample:
\begin{align*}
    \mu &= \frac{1}{N} \sum_{i=1}^N z_i, \\
    \sigma^2 &= \frac{1}{N} \sum_{i=1}^N (z_i - \mu)^2.
\end{align*}
Each activation is then normalized as
\begin{align*}
    \hat{z}_i = \frac{z_i - \mu}{\sqrt{\sigma^2 + c}},
\end{align*}
where $c$ is a small constant to prevent numerical instability. In our experiments, we use $c=10^{-5}$. We omit the additional gain and bias terms often used in LN in our experiments, because existing literature indicates that the gain and bias occasionally hurts training \citep{xu2019understanding}.

\subsection{Network Architecture \& Loss Functions}

We enforce Dale's principle in our networks by constraining each neuron to be exclusively excitatory (E) or inhibitory (I), such that all outgoing synaptic weights share the same sign. Our networks are limited to 2 hidden layers, with the inhibitory layer width set to $10\%$ of the excitatory layer width. The architecture is trained end-to-end using standard gradient descent.

\paragraph{Standard EI Network}

In the standard EI-Network, the activity of the excitatory units ($\mathbf{h}_{\ell}^E$) at layer $\ell$ is governed by the subtractive interaction of E and I populations:

$$
\mathbf{h}_{\ell}^E = \phi(\mathbf{z}_{\ell}) \quad \text{where} \quad \mathbf{z}_{\ell} = \mathbf{W}_{\ell}^{EE} \mathbf{h}_{\ell-1}^E - \mathbf{W}_{\ell-1}^{EI} \mathbf{h}_{\ell}^I.
$$

The inhibitory population activity ($\mathbf{h}_{\ell}^I$) is a feedforward function of the preceding excitatory activity:
$$
\mathbf{h}_{\ell}^I = \mathbf{W}_{\ell}^{IE} \mathbf{h}_{\ell-1}^E.
$$
Here, $\phi(x) = \text{ReLU}(x)$ is the non-linear activation function. For E-only networks (Fig.\ref{fig:fig1}), the inhibitory computation ($\mathbf{h}_{\ell}^I$) is excluded, simplifying the pre-activation to $\mathbf{z}_{\ell} = \mathbf{W}_{\ell}^{EE} \mathbf{h}_{\ell-1}^E$.

\paragraph{The Inhibitory Normalization (I-Norm) Network}

To approximate Layer Normalization (LN) using segregated inhibitory populations, we extended the EI-network architecture by introducing a separate inhibitory population dedicated to divisive inhibition ($\mathbf{h}_{\ell}^D$).

The resulting $\text{I-Norm}$ network equations introduce divisive normalization to the pre-activation $\mathbf{z}_{\ell}$:

$$
\mathbf{h}_{\ell}^E = \phi(\mathbf{z}_{\ell}) \quad \text{where} \quad \mathbf{z}_{\ell} = \frac{\mathbf{W}_{\ell}^{EE} \mathbf{h}_{\ell-1}^E - \mathbf{W}_{\ell-1}^{EI} \mathbf{h}_{\ell}^{I}}{\sqrt{\mathbf{U}_{\ell}^{EI} \mathbf{h}_{\ell}^D}} \quad (\text{Eq. } \ref{eq:inorm})
$$

The subtractive ($\mathbf{h}_{\ell}^I$) and divisive ($\mathbf{h}_{\ell}^D$) inhibitory populations are computed as:
$$
\mathbf{h}_{\ell}^I = \mathbf{W}_{\ell}^{IE} \mathbf{h}_{\ell-1}^E \quad \text{and} \quad \mathbf{h}_{\ell}^D = \mathbf{U}_{\ell}^{IE} \mathbf{h}_{\ell-1}^E.
$$
The weights associated with the divisive pathway are denoted by $\mathbf{U}^{X}$.

\paragraph{Initialization of Standard EI-network}

Following the procedures established by \cite{cornford2021learning}, all excitatory parameters ($\mathbf{W}^{EE}, \mathbf{W}^{IE}$) are initialized from an exponential distribution: $\mathbf{W}^{EE}_{ij} \sim \mathrm{Exp}(\lambda^E)$.

The inhibitory parameters are initialized to ensure that excitation and subtractive inhibition are balanced ($\mathbb{E}[z^E_k] = \mathbb{E}[(\mathbf{W}^{EI} \mathbf{z}^I)_k]$). Specifically:
$\mathbf{W}^{IE}$ is initialized using the mean of the rows of $\mathbf{W}^{EE}$: $\mathbf{W}^{IE} = \frac{1}{n_e} \sum_{j=1}^{n_e} \mathbf{w}^{EE}_{j,:}$.
$\mathbf{W}^{EI}$ is initialized from $\mathrm{Exp}(\lambda^E)$ and then row-normalized ($\mathbf{W}^{EI}_{i,:} \leftarrow \frac{\mathbf{W}^{EI}_{i,:}}{\sum_k \mathbf{W}^{EI}_{ik}}$), which approximates the balancing constant $\frac{1}{n_i}$.

\paragraph{Initialization of I-Norm Divisive Pathway}

To maintain consistency and ensure $\mathbb{E}[\mathbf{z}^E]=0$ at initialization, the subtractive inhibition pathway ($\mathbf{W}^{EI}, \mathbf{W}^{IE}$) is initialized exactly as in the standard EI-network.

For the divisive pathway ($\mathbf{U}^{IE}, \mathbf{U}^{EI}$), the goal is to initialize the denominator ($\sqrt{\mathbf{U}^{EI} \mathbf{h}^D}$) to approximate the empirical variance of the subtractive pre-activations. We achieve this by employing a Singular Value Decomposition (SVD) of the effective excitatory weight matrix $\mathbf{W} = \mathbf{W}^{EX} - \mathbf{W}^{EI} \mathbf{W}^{IX}$ (where $\mathbf{W}^{IX}$ is the $\mathbf{W}^{IE}$ weights):
$$
\mathbf{W} = \mathbf{U} \mathbf{\Sigma} \mathbf{V}^{\top}.
$$
The principal components of $\mathbf{V}$ are used to initialize the divisive pathway:
$$
\mathbf{U}^{IE} = \mathbf{\Sigma} \mathbf{V}^{\top}, \quad \text{and} \quad \mathbf{U}^{EI}_{ij} = \frac{1}{n_e}.
$$
This ensures that the denominator approximates the required empirical variance at the initialization.

We note that the use of SVD for $\mathbf{U}^{IE}$ initialization does not guarantee that all resulting weights are positive, so technically, the divisive pathway in our I-Norm network breaks with Dale's Law. However, we note that divisive inhibition in real neurons is likely driven by shunting  \citep{carandini1994summation}, and shunting inhibition is more likely to be able to switch signs due to chloride dynamics in dendrites \citep{raimondo2012short}.

\paragraph{Training Procedure and Dual Learning Objectives}

The network optimizes excitatory and inhibitory synapses under distinct learning objectives to decouple task learning from neural activity normalization.

Excitatory connections ($\mathbf{W}^{EE}$) are trained on the standard cross-entropy classification loss ($\mathcal{L}_{\text{task}}$), following the rule: $\Delta \mathbf{W}^{EE} = -\eta_{EE} \frac{\partial \mathcal{L}_{\text{task}}}{\partial \mathbf{W}^{EE}}$. Inhibitory connections ($\mathbf{W}^{EI}, \mathbf{W}^{IE}, \mathbf{U}^{EI}, \mathbf{U}^{IE}$) are optimized with a local loss function ($\mathcal{L}_{\text{I-Norm}}$) designed to preserve stability by enforcing the statistics of layer normalization (mean $= 0$, variance $= 1$):
$$
\mathcal{L}_{\text{I-Norm}} = \left ( \frac{1}{n} \sum_{i=1}^n h_i^E\right ) ^2 + \left ( \frac{1}{n} \sum_{i=1}^n {(h_i^E)}^2 -1 \right )^2.
$$

To ensure that stability mechanisms do not interfere with task learning, gradient isolation is enforced using stop-gradient mechanisms:
The excitatory activations are detached when computing the I-Norm loss.
The inhibitory outputs are detached during the main forward pass (propagation of $\mathcal{L}_{\text{task}}$), preventing I-Norm gradients from affecting the task objective.

\subsection{Hyperparameters:}

\begin{table}[h!]
\centering
\caption{Hyperparameters used in the experimental evaluation}
\label{tab:hyperparameters}
\begin{tabular}{@{}lll@{}}
\toprule
\textbf{Parameter} & \textbf{Value/Range} & \textbf{Description} \\
\midrule
\multicolumn{3}{l}{\textbf{Training Parameters}} \\
Epochs & 50 & Number of training epochs \\
Batch size & 32 & Mini-batch size \\
Dataset & FashionMNIST & Source dataset \\
\midrule
\multicolumn{3}{l}{\textbf{Learning Rates}} \\
Excitatory LR ($\eta_{exc}$) & $10^{-3}$ to $10^{-1}$ & Log-uniform sampling \\
Inhibitory LR ($\eta_{wei}$) & $10^{-5}$ to $10^{-2}$ & Log-uniform sampling \\
Inhibitory LR ($\eta_{wix}$) & $10^{-2}$ to $10^{0}$ & Log-uniform sampling \\
\midrule
\multicolumn{3}{l}{\textbf{Network Architecture}} \\
Hidden Layer depth & 2 & Fixed \\
Hidden layer width & 100-500 & Uniform sampling \\
Output classes & 10 & FashionMNIST classes \\
\midrule
\multicolumn{3}{l}{\textbf{I-Norm Loss Weight}} \\
$\lambda_{I-Norm}$ & $10^{-5}$ to $10^{0}$ & Grid Search \\
\midrule
\multicolumn{3}{l}{\textbf{Optimization}} \\
Momentum & 0 & SGD momentum \\
Weight decay & 0 & L2 regularization \\
Algorithm & SGD & Optimizer type \\
\bottomrule
\end{tabular}
\end{table}

We conducted a comprehensive hyperparameter sweep to evaluate the performance on the FashionMNIST dataset. Our experimental design employed a grid search strategy combined with random sampling to explore the hyperparameter space systematically (Table \ref{tab:hyperparameters}).

\textbf{Training Configuration:} All experiments were trained for 50 epochs using mini-batches of size 32. We used the FashionMNIST dataset with 10 output classes. The training employed SGD optimization with no momentum or weight decay to isolate the effects of the I-Norm mechanisms.

\textbf{Learning Rate Sampling:} We implemented separate learning rates for excitatory and inhibitory connections, sampled from log-uniform distributions. The excitatory learning rate ($\eta_{exc}$) was sampled from $[10^{-3}, 10^{-1}]$, while inhibitory learning rates for excitatory-inhibitory ($\eta_{wei}$) and inhibitory-inhibitory ($\eta_{wix}$) connections were sampled from $[10^{-5}, 10^{-2}]$ and $[10^{-2}, 10^{0}]$ respectively. We employed the same inhibitory learning rates for both the subtractive and divisive inhibitory components. This design reflects the biological principle that inhibitory plasticity operates on different timescales than excitatory plasticity.

\textbf{Network Architecture:} The hidden layer width was randomly sampled from a uniform distribution over $[100, 500]$ neurons, allowing us to evaluate the robustness of I-Norm mechanism across different network sizes. We maintained the same hidden layer depth of 2 hidden layers.

\textbf{I-Norm Loss Parameters:} We systematically varied the I-Norm loss weight ($\lambda_{homeo}$) across five values: $10^{-5}$, $10^{-4}$, $10^{-3}$, $10^{-2}$,$10^{-1}$, and $10^{0}$. The default I-norm weight we show in our results is $10^{-2}$.

\textbf{Data Augmentation:} To test robustness to input variations, we applied brightness jitter with factors of 0, 0.25, 0.5, and 0.75, simulating varying lighting conditions.

\subsection{Measures and Analysis}

\paragraph{Cosine Similarity Analysis}

To measure the alignment of the internal dynamics of the $\text{I-Norm}$ network against the explicit $\text{LN}^+$ baseline. We computed the cosine similarity for both the activity outputs and the gradient signals.

We measured the output alignment between the excitatory unit outputs ($\mathbf{h}^{E}_l$) of the $\text{I-Norm}$ network and the $\text{LN}^+$ baseline at layer $l$:
$$
\text{output\_alignment}_l = \frac{(\mathbf{h}^{E}_l)^{\text{I-Norm}} \cdot (\mathbf{h}^{E}_l)^{\text{LN}^+}}{||(\mathbf{h}^{E}_l)^{\text{I-Norm}}||_2 ||(\mathbf{h}^{E}_l)^{\text{LN}^+}||_2}.
$$

We similarly measured the alignment of the gradient signals, specifically the cosine similarity between the gradient of the $\mathcal{L}_{\text{I-Norm}}$ loss and the task-driven gradient of the $\mathcal{L}_{\text{Task}}$ loss, with respect to the excitatory weights $\mathbf{W}^{EE}_l$. Note that the $\mathcal{L}_{\text{Task}}$ loss is computed with respect to the excitatory-only network with $\text{LN}^+$.

$$
\text{gradient\_alignment}_l = \frac{\nabla_{\mathbf{W}^{EE}_l} \mathcal{L}_{\text{I-Norm}} \cdot \nabla_{\mathbf{W}^{EE}_l} \mathcal{L}_{\text{Task}}}{||\nabla_{\mathbf{W}^{EE}_l} \mathcal{L}_{\text{I-Norm}}||_2 ||\nabla_{\mathbf{W}^{EE}_l} \mathcal{L}_{\text{Task}}||_2}.
$$

\paragraph{Statistical Moments of Neural Activations}

To confirm that the $\text{I-Norm}$ mechanism successfully learned to enforce normalization constraints, we tracked the \textbf{first and second moments} of the excitatory pre-activations ($\mathbf{z}_l$) throughout training, where $l$ is the layer.

\begin{itemize}
    \item \textbf{First Moment (Mean):} $\mu_l = \mathbb{E}[\mathbf{z}_l] = \frac{1}{N} \sum_{i=1}^{N} z_{l,i}$.
    \item \textbf{Second Moment:} $\sigma_l^2 = \mathbb{E}[\mathbf{z}_l^2] = \frac{1}{N} \sum_{i=1}^{N} z_{l,i}^2$.
\end{itemize}
The objective of the $\mathcal{L}_{\text{I-Norm}}$ loss is to drive these moments toward the $\text{LN}$ targets (mean $= 0$, variance $= 1$).

\paragraph{Statistical Significance Testing}

To robustly determine statistical significance for comparisons between different network architectures and training conditions (e.g., accuracy comparisons across hyperparameter runs), we employed the \textbf{Mann-Whitney U Test}. This non-parametric test was chosen because the accuracy distributions obtained from hyperparameter sampling may not strictly follow a normal distribution. The results of this test are reported using conventional notations.

\subsection{Implementation and Reproducibility}

All experiments were implemented in PyTorch and conducted on NVIDIA RTX 8000 GPUs with 16GB memory allocation. Training was performed using SLURM job arrays with 30 random hyperparameter configurations per experimental condition, where each run required approximately 20 minutes of compute time. The codebase includes automated batch scripts for hyperparameter sweeps and random configuration generation, ensuring consistent experimental protocols across all runs. The full repository is publicly available at: https://github.com/RoyHEyono/inhibitory-normalization.

\subsection{Acknowledgments}

The authors would like to thank Tom George and Ibrahima Daw for helpful comments on the
manuscript. This work was supported by the following sources RHE: Deepmind fellowship. DL: FRQNT Strategic Clusters Program (2020-RS4-265502 – Centre UNIQUE – Unifying Neuroscience and Artificial Intelligence – Québec), the Richard and Edith Strauss Postdoctoral Fellowship in Medicine and the Thomas Kingsley Lawrence Fund. BR: This work was supported by NSERC (Discovery Grant: RGPIN-2020- 05105; Discovery Accelerator Supplement: RGPAS-2020-00031), CIFAR (Canada AI Chair; Learning in Machine and Brains Fellowship), and DoD OUSD (R\&E) under Cooperative Agreement PHY-2229929 (The NSF AI Institute for Artificial and Natural Intelligence). This research was enabled in part by support provided by (Calcul Québec) (https://www.calculquebec.ca/en/) and the Digital Research Alliance of Canada (https://alliancecan.ca/en). The authors acknowledge the material support of NVIDIA in the form of computational resources.

\newpage
\section{Appendix}

\subsection{Derivative of Layer Normalization}

We begin by establishing the objective function for our network. Let's define the cross entropy
\begin{equation*}
\mathcal{L}_{\text{task}} = - \sum_i y_i \log(\hat{y}_i),
\end{equation*}

where $y_i$ is the label and $\hat{y}_i = softmax(\phi(z^{L}))$ is the prediction of the network. The error at the final softmax layer $L$ can be defined as:

\begin{equation*}
\delta^{L} = \nabla_{\hat{y}}\mathcal{L}_{\text{task}} \odot \phi^{'}(z^{L}),
\end{equation*}

where $\phi$ is the activation function. In the final output layer $L$, $\phi^L$ is defined as the softmax function. For all preceding hidden layers $l < L$, we employ the Rectified Linear Unit (ReLU) activation, $\phi^l(z) = \max(0, z)$. 

Using the chain rule, we can propagate this error backward through the network. The backpropagated error for earlier layers $l$ is defined as:

\begin{equation*}
\delta^{l} = (W^{EE}_{l+1})^{T}\delta^{l+1} \odot \phi^{'}(z^{l}).
\end{equation*}

With the error signal established, we can define the gradient for the weight parameters. The weight update for excitatory weights $W^{EE}_{l+1}$ w.r.t to the error at $\delta^{l+1}$ is defined as:

\begin{equation*}
   \frac{\partial \mathcal{L}_{\text{task}}}{\partial W^{EE}_{l+1}} = \delta^{l+1}(h^E_{l})^{T}.
\end{equation*}

In standard architectures, layer normalization is often introduced before the activation function. Layer normalization is defined as:

\begin{equation}
    \label{eqn:ln_eqn}
    \hat{z}^l = \frac{z^l - \mu^l}{\sigma^l}, 
    \mu^l = \frac{1}{H} \sum_{i=1}^H z^l,
    \sigma^l = \sqrt{\frac{1}{H} \sum_{i=1}^H (z^l - \mu^l)^2 + c}.
\end{equation}

To backpropagate through this operation, we must account for the dependency of the normalized output $\hat{z}^l$ on the input vector $z^l$. When layer normalization is applied to layer $l$, the error signal at layer $l$ becomes:

\begin{equation}
\label{eqn:delta_norm}
\delta_{norm}^{l} = \left( \frac{\partial\hat{z}^{l}}{\partial z^{l}} \right)^T \delta^{l}.
\end{equation}

And hence, the weight update for the preceding layer now incorporates this adjusted error:

\begin{equation*}
   \frac{\partial \mathcal{L}_{\text{task}}}{\partial W^{EE}_{l+1}} = \delta^{l+1}_{norm}(h^E_{l})^{T}.
\end{equation*}

Let's now compute the Jacobian $\frac{\partial \hat{z}^l}{\partial z^l}$ in $\delta^{l}_{norm}$. This represents how each component of the normalized vector changes with respect to each component of the input vector. We first apply the quotient rule to $\frac{\partial \hat{z}^l}{\partial z^l}$ from Equation \ref{eqn:ln_eqn}:

\begin{align*}
    \frac{\partial \hat{z}^l}{\partial z^l} &= \frac{1}{\sigma^l} \frac{\partial}{\partial z^l} \bigg(z^l - \mu^l\bigg) - \frac{z^l - \mu^l}{(\sigma^l)^2} \frac{\partial \sigma^l}{\partial z^l} \\
    &= \frac{1}{\sigma^l} \frac{\partial \big(z^l - \mu^l\big)}{\partial z^l}  - \frac{z^l - \mu^l}{(\sigma^l)^2} \frac{\partial \sigma^l}{\partial z^l} \\
    &= \frac{1}{\sigma^l} \bigg( \boldsymbol{I}_H - \frac{\partial \mu^l}{\partial z^l} - \frac{z^l - \mu^l}{\sigma^l}\frac{\partial \sigma^l}{\partial z^l} \bigg) \\
    &= \frac{1}{\sigma^{l}} \left( \boldsymbol{I}_{H} - \frac{\partial\mu^{l}}{\partial z^{l}} - \hat{z}^{l} \left( \frac{\partial\sigma^{l}}{\partial z^{l}} \right)^T \right).
\end{align*}

To complete the derivation, we solve for the partial derivatives of the mean ($\mu$) and standard deviation ($\sigma$). Note that,
\begin{equation*}
    \frac{\partial \mu^l}{\partial z^l} = \frac{1}{H} \mathbf{1}_H (\mathbf{1}_H)^T,
\end{equation*}

\begin{equation*}
    \frac{\partial \sigma^l}{\partial z^l} = \frac{1}{H} \bigg(\frac{z^l - \mu^l}{\sigma^l}\bigg) = \frac{1}{H} \hat{z}^l.
\end{equation*}

By substituting these intermediate results back into our Jacobian expression, we arrive at the full matrix representation. Altogether,
\begin{align*}
    \frac{\partial \hat{z}^l}{\partial z^l} &= \frac{1}{\sigma^l} \bigg( \boldsymbol{I}_H - \frac{1}{H} \mathbf{1}_H (\mathbf{1}_H)^T - \hat{z}^l\frac{\partial \sigma^l}{\partial z^l} \bigg) \\
    &= \frac{1}{\sigma^l} \bigg( \boldsymbol{I}_H - \frac{1}{H} \mathbf{1}_{H \times H} - \frac{1}{H} (\hat{z}^l) (\hat{z}^l)^T \bigg).
\end{align*}

When we re-introduce Equation \ref{eqn:delta_norm}, the following expression for the normalized error emerges:

\begin{align*}
\delta^{l}_{norm} &= \left( \frac{\partial\hat{z}^{l}}{\partial z^{l}} \right)^T \delta^{l} \\
&= \frac{1}{\sigma^{l}} \left( \mathbf{I}_{H} - \frac{1}{H} \mathbf{1}_{H \times H} - \frac{1}{H} \hat{z}^{l} (\hat{z}^{l})^{T} \right) \delta^{l}.
\end{align*}

For implementation purposes, it is often useful to view the transformation on a per-element basis. Elementwise, this translates to:

\begin{align*}
( \delta^{l}_{\text{norm}})_i
&= \sum_{j=1}^{H} \delta^l_j \frac{\partial \hat{z}^l_j}{\partial z^l_i} \\
&= \frac{1}{\sigma^l} \left( 
\delta^l_i - \frac{1}{H} \sum_{j=1}^{H} \delta^l_j 
- \frac{\hat{z}^l_i}{H} \sum_{j=1}^{H} \delta^l_j \hat{z}^l_j
\right).
\end{align*}

\subsection{Mean estimation via fixed random inhibition}

\textbf{Theorem 1:}

\noindent Let $\{\omega_i\}_{i=1}^{n}$ be i.i.d. positive random variables with $\mathbb{E}[\omega_i] = \mu > 0$ and $\operatorname{Var}(\omega_i) < \infty$ that parameterize a set of $n$ synaptic weights, $\{\nu_{i,n}\}_{i=1}^{n}$, whose values sum to $1$.Let $\{\delta_i\}_{i=1}^{n}$ be a bounded sequence of error signals with an empirical mean denoted by $\bar{\delta}$, i.e.,$$\bar{\delta} = \frac{1}{n}\sum_{i=1}^n \delta_i.$$

Define the normalized synaptic weights$$\nu_{i,n} = \frac{\omega_i}{\sum_{j=1}^n \omega_j},$$and the inhibitory pooling operation, which can be expressed as the dot product between the $n$-dimensional weight vector $\boldsymbol{\nu}_n = [\nu_{1,n}, \dots, \nu_{n,n}]^\top$ and the error vector $\boldsymbol{\delta}_n = [\delta_1, \dots, \delta_n]^\top$:$$s_n = \boldsymbol{\nu}_n \cdot \boldsymbol{\delta}_n = \sum_{i=1}^n \nu_{i,n}\,\delta_i.$$Then, in the limit as $n \to \infty$, the pooled signal converges to the empirical mean:$$\lim_{n \to \infty} s_n = \bar{\delta}. \quad$$

\noindent \textbf{Proof:}

\noindent We first rewrite the pooled signal as a ratio:
\[
s_n
= \sum_{i=1}^n \nu_{i,n}\,\delta_i
= \frac{\sum_{i=1}^n \omega_i \delta_i}{\sum_{j=1}^n \omega_j}.
\]
Dividing numerator and denominator by $n$ gives
\[
s_n
=
\frac{\frac{1}{n}\sum_{i=1}^n \omega_i \delta_i}
     {\frac{1}{n}\sum_{j=1}^n \omega_j}.
\]

\paragraph{Convergence of the denominator.}
By Kolmogorov’s Strong Law of Large Numbers (SLLN) \citep{kolmogoroff1933grundbegriffe},
\[
\frac{1}{n}\sum_{j=1}^n \omega_j \;\to\; \mu
\quad\text{as } n\to\infty.
\]

\paragraph{Convergence of the numerator.}
Decompose:
\[
\frac{1}{n}\sum_{i=1}^n \omega_i \delta_i
=
\frac{1}{n}\sum_{i=1}^n (\omega_i - \mu)\delta_i
\;+\;
\mu\,\frac{1}{n}\sum_{i=1}^n \delta_i.
\]

Because the sequence $\{\delta_i\}$ is bounded, say $|\delta_i| \leq C$, we have
\[
\operatorname{Var}\!\big((\omega_i - \mu)\delta_i\big)
\leq C^2\,\operatorname{Var}(\omega_i)
< \infty.
\]
Thus the random variables $(\omega_i - \mu)\delta_i$ are independent,
mean-zero, and uniformly square-integrable. Kolmogorov’s Strong Law of Large
Numbers for independent, non-identically distributed random variables therefore
implies
\[
\frac{1}{n}\sum_{i=1}^n (\omega_i - \mu)\delta_i \;\to\; 0
\quad
\]

By assumption on $\{\delta_i\}$,
\[
\frac{1}{n}\sum_{i=1}^n \delta_i \;\to\; \bar{\delta},
\]
hence
\[
\frac{1}{n}\sum_{i=1}^n \omega_i \delta_i
\;\to\;
\mu \bar{\delta}.
\quad
\]

\paragraph{Taking the ratio.}
Since the denominator converges to $\mu > 0$, we obtain
\[
s_n
=
\frac{\frac{1}{n}\sum_{i=1}^n \omega_i \delta_i}
     {\frac{1}{n}\sum_{j=1}^n \omega_j}
\;\to\;
\frac{\mu\bar{\delta}}{\mu}
= \bar{\delta}.
\quad
\]

\hfill\textit{(End of Proof 1)}

\newpage

\bibliographystyle{apalike} 
\bibliography{references}   

\end{document}